# SPEAKER RECOGNITION BY MEANS OF A COMBINATION OF LINEAR AND NONLINEAR PREDICTIVE MODELS[1]

*Marcos Faundez-Zanuy*
Escola Universitària Politècnica de Mataró
Universitat Politècnica de Catalunya (UPC)
Avda. Puig i Cadafalch 101-111, E-08303 Mataró (BARCELONA)
e-mail: faundez@tecnocampus.cat

## ABSTRACT

This paper deals the combination of nonlinear predictive models with classical LPCC parameterization for speaker recognition. It is shown that the combination of both a measure defined over LPCC coefficients and a measure defined over predictive analysis residual signal gives rise to an improvement over the classical method that considers only the LPCC coefficients. If the residual signal is obtained from a linear prediction analysis, the improvement is 2.63% (error rate drops from 6.31% to 3.68%) and if it is computed through a nonlinear predictive neural nets based model, the improvement is 3.68%. An efficient algorithm for reducing the computational burden is also proposed.

## 1. INTRODUCTION

In the last years there has been a growing interest for nonlinear models applied to speech. This interest is based on the evidence of nonlinearities in the speech production mechanism. Several arguments justify this fact:
a) Residual signal of predictive analysis [1].
b) Correlation dimension of speech signal [2].
c) Fisiology of the speech production mechanism [3].
d) Probability density functions [4].
e) High order statistics [5].
Although these evidences, few applications have been developed so far. Mainly due to the high computational complexity and difficulty of analyzing the nonlinear systems.
The applications of the nonlinear predictive analysis have been focused on speech coding, because it achieves greater prediction gains than LPC. The most relevant systems are [6] and [7], that have proposed a CELP with different nonlinear predictors that improve the SEGSNR of the decoded signal.
Although the relevance of the LPC residual signal for speech coding (Multi-pulse LPC, CELP, etc.) is well established, little attention has been dedicated to this signal for speaker recognition purposes.
In [8] we proposed and ADPCM with nonlinear prediction based on MLP. Our scheme outperforms the ADPCM with linear prediction of the same order between 1 and 2 dB in SEGSNR.
In [9] we showed the relevance of the residual signal of predictive analysis in speaker recognition.
In [10] it has been shown that humans can recognize the identity of persons by listening to the LPC residual signal, and several authors have used pitch frequency for speaker recognition. On the other hand, it has been found [11] that the residue as a whole carries richer information than the fundamental frequency alone, and the use of a cepstrum computed over the LPC residual signal has been proposed. The use of this parameterization is less efficient than the cepstrum of the LPC coefficients, but a combination of LPC cepstrum and LPC residual cepstrum produces a reduction in the error rate from 5.7% to 4.0%. Although in [11] the residual signal is used, the residue of the LPC residual cepstrum is ignored.

One of the most promising approaches to nonlinear prediction are the neural nets. The main reason of the difficulty for applying the nonlinear predictive models based on MLPs in recognition applications is that it is not possible to compare the nonlinear predictive models directly. The comparison between two predictive models can be done alternatively in the following way: the same input is presented to several models, and the decision is made based on the output of each system, instead of the structural parameters of each system.

For speaker recognition purposes we propose to model each speaker with a codebook of nonlinear predictors based on MLP. This is done in the same way as the classical speaker recognition based on vector quantization [12].

We obtained [9] that the residual signal is less efficient than the LPCC coefficients. Both in linear and nonlinear predictive analysis the recognition errors are around 20%, while the results obtained with LPCC are around 6%. On the other hand we have obtained that the residual signal is uncorrelated with the vocal tract information (LPCC coefficients) [9]. For this reason both measures can be combined in order to improve the recognition rates.

We propose:
a) The use of an error measure defined over the LPC-residual signal, (instead of a parameterization over this signal) combined with a classical measure defined over LPCC coefficients.
b) The use of a nonlinear prediction model based on neural nets, which has been successfully applied to a waveform speech coder [8]. It is well known that the LPC model is unable to describe the nonlinearities present in the speech, so useful information is lost with the LPC model alone.

With a nonlinear prediction model based on neural nets it is not possible to compare the weights of the neural net. This is due to the fact that infinite sets of different weights representing the same model exist, and direct comparison is not feasible. For this reason the measure is defined over the residual signal of the nonlinear prediction model, and so the residual signal is considered in a natural way. For improving performance upon classical methods a combination with linear parameterization must be used. In order to reduce the computational complexity and to improve the recognition rates a novel scheme that consists on the preselection of the K speakers nearest to the test sentence is proposed. Then, the error measure based on the nonlinear predictive model is computed only with these speakers. (In this case a reduction of 3.68% in error rate upon classical LPC cepstrum parameterization is achieved).

[1] This work has been supported by the CICYT TIC97-1001-C02-02

## 2. SPEAKER RECOGNITION USING LPCC COEFFICIENTS AND RESIDUAL SIGNAL

### 2.1 Database

Our experiments have been computed over 38 speakers from the New England dialect of the DARPA TIMIT Database (24 males&14 females). The speech samples were downsampled from 16KHz to 8 KHz, and pre-emphasized by a first order filter whose transfer function was $H(z)=1-0.95z^{-1}$. A 30ms Hamming window was used, and the overlapping between adjacent frames was 2/3. A cepstral vector of order 12 was computed from the LPC coefficients. Five sentences are used for training, and 5 sentences for testing (each sentence is between 0.9 and 2.8 seconds long).

### 2.2 Recognition algorithm baseline

Our recognition algorithm is a Vector Quantization approach. That is, each speaker is modelled with a codebook in the training process. During the test, the input sentence is quantized with all the codebooks, and the codebook which yields the minimal accumulated error indicates the recognized speaker.

The codebooks are generated with the splitting algorithm. Two methods have been tested for splitting the centroids:
a) The standard deviation of the vectors assigned to each cluster.
b) A hyperplane computed with the covariance matrix.
The recognition algorithm in the MLP's codebook is the following:
   1. The test sentence is partitioned into frames.For each speaker:
   2. Each frame is filtered with all the MLP of the codebook (centroids) and it is stored the lowest Mean Absolute Error (MAE) of the residual signal. This process is repeated for all the frames, and the MAE of each frame is accumulated for obtaining the MAE of the whole sentence.
   3. The step 2 is repeated for all the speakers, and the speaker that gives the lowest accumulated MAE is selected as the recognized speaker.

This procedure is based on the assumption that if the model has been derived from the same speaker of the test sentence then the residual signal of the predictive analysis will be lower than for a different speaker not modeled during the training process.
Unfortunately the results obtained with this system were not good enough. Even with a computation of a generalization of the Lloyd iteration for improving the codebook of MLP. We believe that with additional research this results can be improved. Main subjects that must be studied are:
- An algorithm for clustering the frames. Our studies reveal that this is a fundamental question, with a nontrivial solution.
- Number of epochs and initializations for training the MLP.

Unfortunately the computational burden of the training process is very high and it is difficult to study these points. The simulations were done with a pentium II 266MHz with win NT and C code. In this situation, 4 days are required for computing the 4, 5 and 6 bits codebooks for each speaker (38x3 codebooks), and for one iteration of the generalized Lloyd algorithm.

Figure 1 shows the results obtained with a classical LPCC codebook. Our results compare favorably with Farrell ones [13], because we use the 38 speakers of the New England dialect whereas he uses only 20. This is because our VQ implementation has been more depurated, using the splitting algorithm, and the Hyperplane method for splitting the centroids. Figure 1 also shows the result of a linear combination between the residual signal of LPC analysis with the measure defined over the LPCC coefficients.

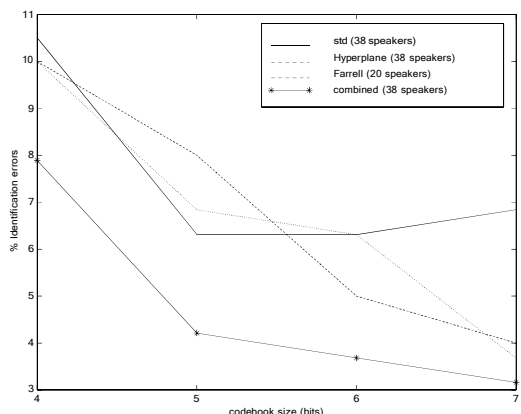

Figure 1. Identification errors for different schemes.

These results can be improved using a combination with the residual signal of the nonlinear predictive analysis, in the same way as they were improved in [9] with the residual signal of LPC analysis. The distortion information obtained from the LPCC coefficients is combined with the error measure defined over the residue, with the following expression:

$error = LPCC\_error + \alpha * residue\_error$ , where the

combination factor has been determined experimentally, by a trial and error procedure, and its value has no special meaning because the combined terms have different origins: the LPCC errors are obtained computing the difference between vectors of dimension 12, and each component has a small value. The residue error is obtained over 240 samples, and each sample has greater magnitude than the LPCC coefficients.

### 2.3 Nonlinear codebook generation

In order to generate the codebook a good initialization must be achieved. Thus, it is important to achieve a good clustering of the train vectors. We have evaluated several possibilities, and the best one is to implement first a linear LPCC codebook of the same size. This codebook is used for clustering the input frames, and then each cluster is the training set for a multilayer perceptron with 10 input neurons, 4 neurons in the first hidden layer, 2 neurons in the second hidden layer, and one output neuron with a linear transfer function. Thus, the MLP is trained in the same way as in [3], but the frames have been clustered previously with a linear LPCC codebook. After this process, the codebook can be improved with a generalization of the Lloyd iteration: frames are clustered with the nonlinear codebook.
   a) new neural nets are trained with the new clusters.
In this process, the weights and biases of the network are initialized with a multi-start algorithm, which consists in training4 different random initializations and the weights of the previous iteration (5 different initializations in total). After the training process, we choose the initialization that gives the lowest quantization distortion. For each initialization 8 epochs are done with the Levenberg-Marquardt algorithm. We tried to use a codebook of LPC coefficients instead of LPCC, but empty cells appeared in the codebooks, yielding worse results.

### 2.4 An efficient recognition algorithm

The application of a codebook of nonlinear predictive models based on MLP gives poor results. Also the computational complexity of the recognition process is high. For these reasons, we propose an efficient algorithm that improves the results of the LPCC codebook alone and that preserves the computational complexity moderate.
The LPCC used for clustering the frames is used as a pre-selectorof the recognized speaker. That is, the input sentence is quantized with the LPCC codebooks and the K codebooks that produce the lowest accumulated error are selected. Then, the input sentence is quantized with the K nonlinear codebooks, and the accumulated distance of the nonlinear codebook is combined with the LPCC distance. With this system, for K=2 the results are (table 1):

The computational complexity of filtering each frame with the neural nets is higher than the comparison of the LPCC coefficients. Thus this scheme implies a reduction of the required number of flops, because the input frames are not filtered with all the codebooks.

We will ignore the computational burden of LPCC computation (that is necessary for all the methods) and will take the following assumptions:

N : number of speakers.
P : predictive analysis order.
$T_{cl}$ : Size of the LPCC codebook.
$T_{cnl}$ : Size of the MLP codebook.
K : number of preselected speakers, using the LPCC codebook
$l_t$ : Frame length.
$n_i$ : Number of neurons in the input layer.
$C_{tg}$ : Number of instructions for the nonlinear transfer function computation.

In this situation the computational burden is proportional to:
For the LPCC codebook:
$$N_{LPCC} = T_{cl} \times p \times N \quad (1)$$
For the MLP codebook:
$$N_{MLP} = N_{LPCC} + K \times T_{cnl} \times (l_t - n_i) \times \left(n_i \times n_{h1} + n_{h1} + n_{h1} \times n_{h2} + 2n_{h2} + 1 + C_{tg} \times (n_{h1} + n_{h2})\right) \quad (2)$$

With a proper substitution and for K=2 we obtain the lowest computational burden of the combined system:
$$N_{LPCC} = 1536N \quad (3)$$
$$N_{MLP} = 1536N + 1.6E6 \quad (4)$$

Obviously if the number of speakers (N) is high, the computational complexity of this system is considerably smaller than for K=N.

This system outperforms the scheme based on LPCC and the combined scheme with linear residue of figure 1. The most relevant conclusions are:
- There is a reduction in error rate of 1% compared with the LPCC codebook of 7 bits, and 3% with respect to 4,5 and 6 bits LPCC codebooks.
- There is a reduction in error rate of 0.5% compared with the best combination between linear residue and LPCC coefficients, for a codebook of 7 bits. For codebooks of 4, 5 and 6 bits the improvement is about 1%.

It is interesting to see that the use of the residue in the nonlinear predictive model implies the use of the residual information, and also of the nonlinear predictive model, because it affects the magnitude of the residue.

| Linear Code-book | Iteration 0 MLP Codebook | | | Iteration 3 MLP Codebook | | |
|---|---|---|---|---|---|---|
| | 4 | 5 | 6 | 4 | 5 | 6 |
| 4 bits | 7.89 | 7.37 | 7.89 | 7.4 | 6.84 | 7.89 |
| 5 bits | 3.68 | 3.16 | 4.74 | 3.7 | 4.21 | 4.74 |
| 6 bits | 3.68 | 3.16 | 4.21 | 3.7 | 3.68 | 3.68 |
| 7 bits | 3.16 | 2.63 | 2.63 | 3.2 | 2.63 | 2.63 |

**Table 1:** Identification errors (%) for different codebook sizes (between 4 and 6 bits) and iterations 0 and 3, for combined scheme.

## 2.5 About α and K selection

The combination factor α has been selected evaluating the optimal value in a sufficiently wide range of values. Figure 2 shows the identification errors as a function of α. It has been obtained for a linear combination between the MAE of the LPCC coefficients and the MSE of the LPC residue on the top, and with the MAE of the LPC residue on the bottom. Thousand different values of have been tested. For the linear combination between the MAE of the LPCC coefficients and the MLP residue the behaviour is very similar. The main conclusion is that the selection of is not a critical subject.

It is more interesting the significance of the K value. Figure 3 shows the identification errors as function of K in a pure nonlinear MLP-VQ recognition scheme. That is, the LPCC codebook used for clustering the frames in the training process is used as a preselector of the recognized speaker. The input sentence is quantized with the LPCC codebooks and the K codebooks that produce the lowest accumulated error are selected. Then, the input sentence is quantized with the K nonlinear codebooks, and the accumulated distance of the nonlinear codebook is selected as the error criterion. The meaning of the extreme values of K are:
- For K=1 the system is equivalent to the LPCC codebook alone. That is, the MLP codebooks have no relevance.
- For K=N the preselector has no relevance.

As previously stated the behaviour of the speaker recognition based on the residual signal of nonlinear predictive analysis does not produce good enough results, even with the use of a preselector of the most probably speakers. On the other hand, if the residual signal of nonlinear predictive analysis is combined with the LPCC information, then the recognition rates are significatively improved upon K=1. Even more than with a combination with the LPC analysis residual signal. If the α combination factor is properly computed the recognition rate is insensitive to K (for 2≤K≤N). Therefore we have chosen the value that gives the lowest computational value: K=2.

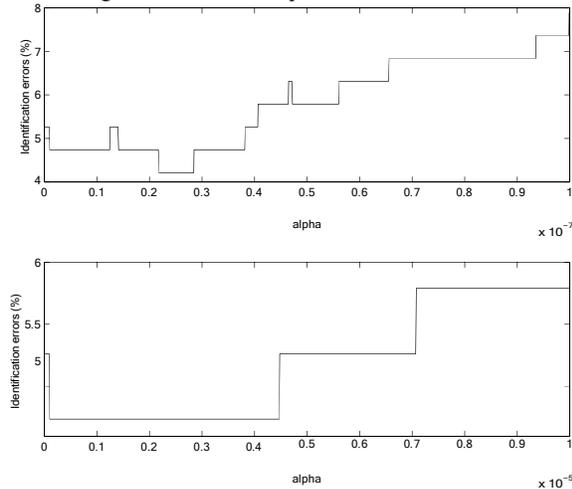

Figure 2. Identification errors versus alpha values for different combinations.

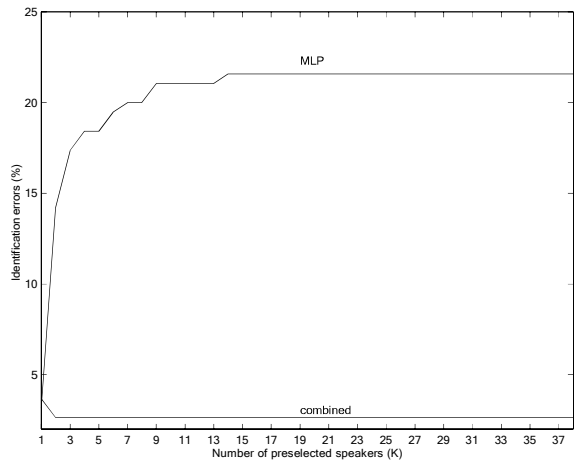

Figure 3. Identification errors for different schemes

## 3. CONCLUSIONS

In this paper we have evaluated the usefulness of nonlinear predictive models based on MLP for speaker recognition. We have found that they must be combined with LPCC parameterization for improving the results of classical LPCC alone. Also we have proposed an efficient algorithm for reducing the computational complexity of the recognizer based on nonlinear predictive models.
Additional research must be done for improving the codebook of nonlinear predictors, and thus to obtain better recognition rates, because better results can be obtained with a combination between other methods [14].

## 4. REFERENCES

[1] J. Thyssen, H. Nielsen y S.D. Hansen "Non-linear short-term prediction in speech coding". ICASSP-94, pp.I-185 , I-188.

[2] B. Townshend "Nonlinear prediction of speech". ICASSP-91, Vol. 1, pp.425-428.

[3] H.M. Teager "Some observations on oral air flow vocalization" IEEE trans. ASSP, vol.82 pp.559-601, October 1980

[4] G. Kubin "nonlinear processing of speech" chapter 16 of Speech coding and synthesis, editors W.B. Kleijn & K.K. Paliwal, Ed. Elsevier 1995.

[5] J. Thyssen, H. Nielsen y S.D. Hansen "Non-linearities in speech",proceedings IEEE workshop Nonlinear Signal & Image Processing, NSIP'95,June 1995

[6] A. Kumar & A. Gersho "LD-CELP speech coding with nonlinear prediction". IEEE Signal Processing letters Vol. 4 Nº4, April 1997, pp.89-91

[7] L. Wu, M. Niranjan & F. Fallside "Fully vector quantized neural network-based code-excited nonlinear predictive speech coding". IEEE transactions on speech and audio processing, Vol.2 nº 4, October 1994.

[8] M. Faundez-Zanuy, Francesc Vallverdu & Enric Monte , "Nonlinearprediction with neural nets in ADPCM" ICASSP-98 .SP11.3 Seattle, USA

[9] M. Faundez-Zanuy & D. Rodríguez "speaker recognition using residual signal of linear and nonlinear prediction models". Vol. 2 pp. 121-124 ICSLP'98. Dec. 1998

[10] T. C. Feustel, G. A. Velius & R. J. Logan "Human and machine performance on speaker identity verification". Speech Tech. 89 pp. 169-170

[11] P. Thevenaz & H. Hugli, "Usefulness of the LPC-residue in text-independent speaker verification". Speech Communication, Vol. 17, pp. 145-157. 1995. Ed. Elsevier.

[12] F. K. Soong, A. E. Rosenberg, L. R. Rabiner y B. H. Juang " A vector quantization approach to speaker recognition". pp. 387-390. ICASSP 1985

[13] K. R. Farrell, R. J. Mammone, K. T. Assaleh. "Speaker recognition using neural networks and conventional classifiers". IEEE Transactions on speech and audio processing. Vol. 2 nº 1, part II, pp.194-205, January 1994.

[14] D. Rodríguez & M. Faundez-Zanuy "Speaker recognition with a MLP classifier and LPCC Codebook". ICASSP'99 NN-SP1.3 Phoenix, USA